\documentclass{aastex631}

\usepackage[T1]{fontenc}
\usepackage{booktabs}
\usepackage{amsmath}

\begin{document}

\title{Gaia21bja: pre-main sequence star with quasi-periodic bursts}

\author{Ádám Mátéfy}
\affiliation{Nagoya University, School of Science,
Department of Physics,
G30 Physics Program, 1 Furo-cho, Chikusa-ku, Nagoya, Aichi, 464-8602, Japan}
\email{matefy.adam.s7@s.mail.nagoya-u.ac.jp}

\author[0000-0002-3632-1194]{Zs\'ofia Nagy}
\affiliation{Konkoly Observatory, HUN-REN Research Centre for Astronomy and Earth Sciences \\ Konkoly-Thege Mikl\'os \'ut 15-17, H-1121 Budapest, Hungary}
\affiliation{CSFK, MTA Centre of Excellence, Konkoly-Thege Mikl\'os \'ut 15-17, 1121 Budapest, Hungary}
\email{nagy.zsofia@csfk.org}

\author[0000-0001-7157-6275]{\'Agnes K\'osp\'al}
\affiliation{Konkoly Observatory, HUN-REN Research Centre for Astronomy and Earth Sciences \\ Konkoly-Thege Mikl\'os \'ut 15-17, H-1121 Budapest, Hungary}
\affiliation{CSFK, MTA Centre of Excellence, Konkoly-Thege Mikl\'os \'ut 15-17, 1121 Budapest, Hungary}
\affiliation{ELTE E\"otv\"os Lor\'and University, Institute of Physics and Astronomy, P\'azm\'any P\'eter s\'et\'any 1A, Budapest 1117, Hungary}

\author[0000-0001-6015-646X]{P\'eter \'Abrah\'am}
\affiliation{Konkoly Observatory, HUN-REN Research Centre for Astronomy and Earth Sciences \\ Konkoly-Thege Mikl\'os \'ut 15-17, H-1121 Budapest, Hungary}
\affiliation{CSFK, MTA Centre of Excellence, Konkoly-Thege Mikl\'os \'ut 15-17, 1121 Budapest, Hungary}
\affiliation{ELTE E\"otv\"os Lor\'and University, Institute of Physics and Astronomy, P\'azm\'any P\'eter s\'et\'any 1A, Budapest 1117, Hungary}
\affiliation{Department of Astrophysics, University of Vienna, Türkenschanzstr. 17, 1180, Vienna, Austria}

\author[0000-0002-4283-2185]{Fernando Cruz-S\'aenz de Miera}
\affiliation{Institut de Recherche en Astrophysique et Plan\'etologie, Universit\'e de Toulouse, UT3-PS, OMP, CNRS, 9 av. du Colonel Roche, 31028 Toulouse Cedex 4, France}
\affiliation{Konkoly Observatory, HUN-REN Research Centre for Astronomy and Earth Sciences \\ Konkoly-Thege Mikl\'os \'ut 15-17, H-1121 Budapest, Hungary}
\affiliation{CSFK, MTA Centre of Excellence, Konkoly-Thege Mikl\'os \'ut 15-17, 1121 Budapest, Hungary}

\author[0000-0002-3648-433X]{M\'at\'e Szil\'agyi}
\affiliation{Konkoly Observatory, HUN-REN Research Centre for Astronomy and Earth Sciences \\ Konkoly-Thege Mikl\'os \'ut 15-17, H-1121 Budapest, Hungary}
\affiliation{CSFK, MTA Centre of Excellence, Konkoly-Thege Mikl\'os \'ut 15-17, 1121 Budapest, Hungary}

\author[0000-0001-5018-3560]{Micha\l\ Siwak}
\affiliation{Mt. Suhora Astronomical Observatory, University of the National Education Commission, ul. Podchora\.zych 2, 30-084 Krak{\'o}w, Poland}
\affiliation{Konkoly Observatory, HUN-REN Research Centre for Astronomy and Earth Sciences \\ Konkoly-Thege Mikl\'os \'ut 15-17, H-1121 Budapest, Hungary}
\affiliation{CSFK, MTA Centre of Excellence, Konkoly-Thege Mikl\'os \'ut 15-17, 1121 Budapest, Hungary}

\author[0000-0002-5261-6216]{Eleonora Fiorellino}
\affiliation{Alma Mater Studiorum – Università di Bologna, Dipartimento di Fisica e Astronomia “Augusto Righi”, Via Gobetti 93/2, I-40129, Bologna, Italy}
\affiliation{INAF - Osservatorio Astronomico di Trieste, via Tiepolo 11, I-34143 Trieste, Italy}

\author[0000-0002-7035-8513]{Teresa Giannini}
\affiliation{INAF-Osservatorio Astronomico di Roma, via di Frascati 33, 00078, Monte Porzio Catone, Italy}

\author[0000-0002-7538-5166]{M\'aria Kun}
\affiliation{Konkoly Observatory, HUN-REN Research Centre for Astronomy and Earth Sciences \\ Konkoly-Thege Mikl\'os \'ut 15-17, H-1121 Budapest, Hungary}
\affiliation{CSFK, MTA Centre of Excellence, Konkoly-Thege Mikl\'os \'ut 15-17, 1121 Budapest, Hungary}

\author[0000-0002-2046-4131]{L\'aszl\'o Szabados}
\affiliation{Konkoly Observatory, HUN-REN Research Centre for Astronomy and Earth Sciences \\ Konkoly-Thege Mikl\'os \'ut 15-17, H-1121 Budapest, Hungary}
\affiliation{CSFK, MTA Centre of Excellence, Konkoly-Thege Mikl\'os \'ut 15-17, 1121 Budapest, Hungary}

\author[0000-0002-1326-1686]{G\'abor Marton}
\affiliation{Konkoly Observatory, HUN-REN Research Centre for Astronomy and Earth Sciences \\ Konkoly-Thege Mikl\'os \'ut 15-17, H-1121 Budapest, Hungary}
\affiliation{CSFK, MTA Centre of Excellence, Konkoly-Thege Mikl\'os \'ut 15-17, 1121 Budapest, Hungary}

\author{Patrik Németh}
\affiliation{ELTE E\"otv\"os Lor\'and University, Institute of Physics and Astronomy, P\'azm\'any P\'eter s\'et\'any 1A, Budapest 1117, Hungary}
\affiliation{Konkoly Observatory, HUN-REN Research Centre for Astronomy and Earth Sciences \\ Konkoly-Thege Mikl\'os \'ut 15-17, H-1121 Budapest, Hungary}
\affiliation{CSFK, MTA Centre of Excellence, Konkoly-Thege Mikl\'os \'ut 15-17, 1121 Budapest, Hungary}

\author[0000-0002-9190-0113]{Brunella Nisini}
\affiliation{INAF-Osservatorio Astronomico di Roma, via di Frascati 33, 00078, Monte Porzio Catone, Italy}

\author[0000-0001-9830-3509]{Zs\'ofia Marianna Szab\'o}
\affiliation{Max Planck Institute for Radio Astronomy, Auf dem Hügel 69, 53121 Bonn, Germany}
\affiliation{Scottish Universities Physics Alliance (SUPA), School of Physics and Astronomy, University of St Andrews, North Haugh, St Andrews, KY16 9SS, UK}
\affiliation{Konkoly Observatory, HUN-REN Research Centre for Astronomy and Earth Sciences \\ Konkoly-Thege Mikl\'os \'ut 15-17, H-1121 Budapest, Hungary}
\affiliation{CSFK, MTA Centre of Excellence, Konkoly-Thege Mikl\'os \'ut 15-17, 1121 Budapest, Hungary}



\begin{abstract}
Gaia21bja is a Gaia alerted young stellar object (YSO) that exhibits at least seven quasi-peridoic brightenings over a 20 year-long light curve with durations of 1.5-2 years and amplitudes up to $\sim$1.7 mag in the Gaia $G$-band.
We analyze its optical and near-infrared photometry and spectra taken using the IRTF and VLT in its faint and bright states in order to characterize its physical properties.
A Lomb-Scargle periodogram analysis results in a most significant period of $916\pm77$ days. We derived the stellar parameters as $R_\star= 0.78 \pm 0.04~R_\odot$, $L_\star=(4.5\pm0.3) \times 10^{-2}~L_\odot$, and $M_\star= 0.16 \pm 0.03~M_\odot$.
The spectra taken during the burst are dominated by emission lines and are similar to those of EX Lupi-type eruptive young stars (EXors). 
We found that the accretion luminosity and mass accretion rate increased by a factor of $5.5-6$ during the burst. Based on this, and the quasi-periodic bursts, we suggest that Gaia21bja is an eruptive YSO, and is most consistent with the `Periodic' category of the Outbursting YSOs Catalogue.
\end{abstract}



\section{Introduction}
\label{sec:intro}

Young stellar objects (YSOs) show variability due to various physical processes, such as accretion and circumstellar extinction \citep{Fischer2023}. The most important process is the former, which can produce variability with a wide range of time scales and amplitudes. Most YSOs show variability with typical time scales from hours to days and with amplitudes below 1 mag. 
Longer episodes with increased accretion rates also occur and were proposed to explain the observed luminosities of YSOs, which were found to be lower than the expectations \citep{Kenyon1990}. Such events include bursts and outbursts as recently defined by \citet{Fischer2023}. 
Bursts typically occur on time scales from months to a year with amplitudes of 1--2.5 mag. Outbursts produce even larger amplitudes, $\sim$2.5--6 mag on time scales from months to years to even a century.

Young stellar objects that experience episodic accretion are called eruptive young stars, and are traditionally classified into two main groups: EX Lupi-type stars (EXors, \citealp{Herbig2008}) and FU Orionis-type stars (FUors, \citealp{Herbig1977}). EXors produce bursts or outbursts that last between a month up to a year, some of which are repeated non-periodically every few years, while FUors are in outburst for multiple years or decades. In addition to their time scale, EXors produce bursts/outbursts with typical mass accretion rates of a few $10^{-7}$ $M_\odot$ yr$^{-1}$ (\citealp{ContrerasPena2025} and references therein), while FUor outbursts can have mass accretion rates in excess of $10^{-4}$ $M_\odot$ yr$^{-1}$ (\citealp{Fischer2023} and references therein). Their spectra also show key differences: EXors display rich emission-line spectra, similar to classical T Tauri type stars (CTTS), whereas FUor spectra are dominated by absorption lines.
The optical spectra of EXors are dominated by emission lines thought to be produced in magnetospheres and accretion shocks while the optical-near IR spectra of FUors are dominated by absorption lines produced in the accretion disk (\citealp{Audard2014}, \citealp{Fischer2023}, \citealp{Liu2022}).

A decade ago, only about 40 eruptive young stars were known \citep{Audard2014}. Currently there are 173 confirmed eruptive YSOs and 355 candidates, according to the new catalog of outbursting YSOs (OYCAT, \citealp{ContrerasPena2025})\footnote{\href{http://starformation.synology.me:5002/OYCAT/main.html}{http://starformation.synology.me:5002/OYCAT/main.html}}. This increase is due to transient surveys including the Gaia Photometric Science Alerts \citep{Hodgkin2021} with its 4$\pi$ sky coverage and approximately monthly cadence.
Several discoveries of eruptive young stars have been made using the Gaia Alerts in recent years, such as four FUors: Gaia17bpi \citep{Hillenbrand2018}, Gaia18dvy \citep{SzegediElek2020}, Gaia21elv \citep{Nagy2023}, and Gaia20bdk \citep{Siwak2025}. Other objects, such as Gaia20eae \citep{CruzSaenzdeMiera2022, Ghosh2022}, Gaia19fct \citep{Park2022}, and Gaia23bab \citep{Giannini2024,Nagy2025}, Gaia21bkw \citep{Giannini2026}, and Gaia24beh \citep{Giannini2026} were identified as EXors. 
Some Gaia alerted eruptive YSOs showed both FUor and EXor properties, such as Gaia19ajj \citep{Hillenbrand2019}, Gaia19bey \citep{Hodapp2020}, Gaia21bty \citep{Siwak2023}, Gaia18cjb \citep{Fiorellino2024}, and Gaia20dsk \citep{Nemeth2026}. 
Another young star, Gaia20fgx, showed brightness variations similar to EXors, however, its mass accretion rate was found to be closer to those of CTTS rather than the values measured for eruptive YSOs \citep{Nagy2022}.
\citet{Giannini2026} found, based on a sample of 16 Gaia alerted YSOs, that those which exhibit variability with amplitudes of $\Delta G \gtrsim 2$ mag and $W1 \gtrsim 1$ mag are likely EXors. 

However, episodic accretion can manifest in smaller scale bursts too, such as the `bursters' reported by \citet{Findeisen2013}, \citet{CodyHillenbrand2018}, and \citet{LakelandNaylor2022}. Bursters occur on shorter time scales compared to eruptive YSOs, i.e., from days-months and with typical amplitudes of 1-2 mag in the $R$-band \citep{Findeisen2013}.
The mass accretion rate and its variability pattern is also expected to change with the evolutionary state of the YSO. For instance, variability amplitudes were shown to be larger for more embedded sources (\citealp{Fischer2023} and references therein).

Gaia21bja (2MASS J16041416-2129151, WISE J160414.16-212915.5, Gaia DR3 6243394091902073728) had a Gaia alert on 2021 February 11 due to a brightening by about 1.5 mag. It is also included in the alerts of the Zwicky Transient Facility (ZTF, \citealp{Patterson2019}) as ZTF19aauxhac. It is located at $\alpha_{J2000}=16^h 04^m 14.17^s,~ \delta_{J2000}=-21^\circ 29' 15.28''$ at a distance of $d = 146.5 \pm 1.3$ pc \citep{BailerJones2021}. 
\cite{Esplin_2018} found that it has a disk that is optically thick at infrared wavelengths and has not yet significantly cleared primordial dust and gas, indicating that Gaia21bja is a YSO. \cite{Luhman_2018} determined its spectral type as M4, corresponding to an effective temperature of $3190\pm75$ K \citep{Thanathibodee2022}. \cite{Luhman_2020} estimated that its $K$ band extinction is $A_K=0.051$ mag, based on the $J-H$ color excess, and have also confirmed it to be a member of the Upper Scorpius star-forming region (Upper Sco). \citet{Ratzenboeck2023a,Ratzenboeck2023b} also confirmed Gaia21bja to belong to Upper Sco. 

In this paper, we analyze the physical properties of Gaia21bja and investigate whether it can be classified as an eruptive YSO. 
In Section \ref{sect_observations} we summarize the available photometric and spectroscopic observations. In Section \ref{sect_results} we determine the stellar and accretion parameters of the star. We discuss our results in Section \ref{sect_discussion}.

\section{Observations and archival data}   
\label{sect_observations}

\subsection{Photometry}

We downloaded $G$ band photometry from the \textit{Gaia} Science Alerts Index website.
We also downloaded the $o$ (“orange”, 560–820 nm) and $c$ (“cyan”, 420–650 nm) band magnitudes from the Asteroid Terrestrial-impact Last Alert System (ATLAS, \citealp{Tonry2018}, \citealp{Smith2020}, \citealp{Heinze2018}) survey using the ATLAS Forced Photometry web service \citep{Shingles2021}.

We used mid-infrared photometry from the Wide-field Infrared Survey Explorer (\textit{WISE}, \citealp{Wright2010}, \citealp{irsa1}) and \textit{NEOWISE} (\citealp{Mainzer2011}, \citealp{irsa144}) surveys from the NASA/IPAC Infrared Science Archive. \textit{NEOWISE} observed the full sky on average twice a year with multiple exposures per epoch. For a comparison with the photometry from other instruments, we computed the average of multiple exposures of a single epoch.
\textit{NEOWISE} $W1$ and $W2$ photometry is known to display a photometric bias for saturated sources. We corrected for this bias using the correction curves given in the Explanatory Supplement to the \textit{NEOWISE} Data Release Products \citep{Cutri2015}.
We derived the average of the uncertainties of the single exposures (err1). We also calculated the standard deviation of the points we averaged per season (err2). For the error of the data points averaged per epoch we used the maximum of err1 and err2.	

We also used photometry from the Catalina Real-Time Transient Survey (CRTS, \citealp{Drake2009}). We obtained $V$-band photometry from the Catalina Surveys Data Release DR2\footnote{http://nesssi.cacr.caltech.edu/DataRelease}. The CRTS covered $\sim$33,000 square degrees and provided light curves for about 500 million objects with $V$ magnitudes between 11.5 mag and 21.5 mag.

In addition to the archival photometry, we took $JHK_s$ images of Gaia21bja with the infrared guiding camera of SpeX, a medium-resolution infrared spectrograph \citep{Rayner2003} on the NASA Infrared Telescope Facility (IRTF) on Mauna Kea, Hawai'i (USA) on 22 July 2023 (Program ID: 2023A968; PI: Cruz S\'aenz de Miera), on 10 August 2024 (Program ID: 2024B025, PI: Siwak), and on 2 March 2025 (Program ID: 2025A019, PI: Kóspál) using a 5-point dither pattern and 5\,s exposures. We used the dithering to make a sky image which we subtracted from the shifted and co-added images. Finally we obtained aperture photometry for Gaia21bja and several other stars in the approximately $1'\times1'$ field of view as comparison stars. We used the Two Micron All Sky Survey (2MASS, \citealp{irsa2}) for the photometric calibration. The resulting magnitudes are $J=11.692\pm0.020$\,mag, $H=10.767\pm0.018$\,mag, $K_s=10.065\pm0.018$\,mag on 22 July 2023, $J=11.977\pm0.079$\,mag, $H=11.360\pm0.016$\,mag, $K_s=10.909\pm0.069$\,mag on 10 August 2024, and $J=12.022\pm0.058$\,mag, $H=11.355\pm0.023$\,mag, $K_s=10.911\pm0.068$\,mag on 2 March 2025.

\subsection{Spectroscopy}

We obtained optical-infrared medium resolution spectra of Gaia21bja using the SpeX spectrograph \citep{Rayner2003} at the 3.2\,m diameter NASA Infrared Telescope Facility (IRTF) located on Mauna Kea, Hawai'i (USA) on 22 July 2023, 10 August 2024, and 2 March 2025. In all cases, we used the SXD grating covering the $0.7-2.55\,\mu$m range and the 0$\farcs$8 wide slit, providing a spectral resolution of about 750. On 2 March 2025, we additionally used the LXD\_long grating that covers $1.98-5.3\,\mu$m range and the 0$\farcs$8 wide slit, providing a spectral resolution of about 940. Total exposure times ranged between 1079\,s and 2158\,s with the SXD grating and 389\,s with the LXD\_long grating. As telluric standards, we observed the A0 stars HD\,133466 (on 22 July 2023) and HD\,145127 (on 10 August 2024 and 2 March 2025) using the same instrumental setup as for the science target. We reduced all spectra in the same way using SpexTool \citep{Vacca2003,Cushing2004}. Data reduction steps included flat correction, subtraction of A and B nod positions, calculation of the spectroscopic traces on the 2D images, extraction of the 1D spectra in tapered apertures, division by the telluric spectra to correct for instrumental response and telluric absorption and emission, and the stitching of the various \'echelle orders. The telluric spectra were also used for a first estimate of the absolute flux calibration. We also derived synthetic photometry by multiplying the flux-calibrated spectra with the respective filter profiles, summing up the obtained values, and converting them from Jy to magnitude. 
We then scaled the spectra so that the synthetic photometry matches the photometry obtained with the guide camera. The scaling factor was 0.82 on 22 July 2023, 1.02 on 10 August 2024, and 1.02 on 2 March 2025, indicating that flux calibration with the telluric standard and the photometric comparison star give consistent results within 20\% or better.

We obtained another spectrum on 9 February 2024 using the X-shooter instrument \citep{Vernet2011}, an
echelle spectrograph mounted on the ESO Very Large Telescope (VLT) at the Paranal Observatory in Chile (Program ID: 112.25XA; PI: Cruz-Sáenz de Miera). 
Covering the spectral range from 300 to 2500 nm, this instrument can simultaneously observe the ultraviolet (UVB), visible (VIS), and near-infrared (NIR) spectra. We used the $0.5'' \times 11''$, $0.4'' \times 11''$, and $0.4'' \times 11''$ size slits providing spectral resolutions of 9700, 18400, and 11600 in the UVB, VIS, and NIR arms, respectively. The exposure time for UVB, VIS, and NIR regions were 1492 s, 1536 s, 1600 s, respectively. We also measured Gaia21bja using the $5'' \times 11''$ slit to correct for slit losses and estimate the actual flux level by scaling the narrow slit spectrum to match the broad slit spectrum. The data were reduced using the X-shooter pipeline \citep{Modigliani2010}, and corrected for tellurics using MOLECFIT \citep{Smette2015,Kausch2015}. We calculated synthetic photometry based on the X-shooter spectrum, which resulted in $J = 11.604$ mag, $H = 10.931$ mag, and $K = 10.422$ mag.

\section{Results}
\label{sect_results}

\subsection{Light and color variations}

\begin{figure}[h!]
\centering
\includegraphics[width=8.9cm]{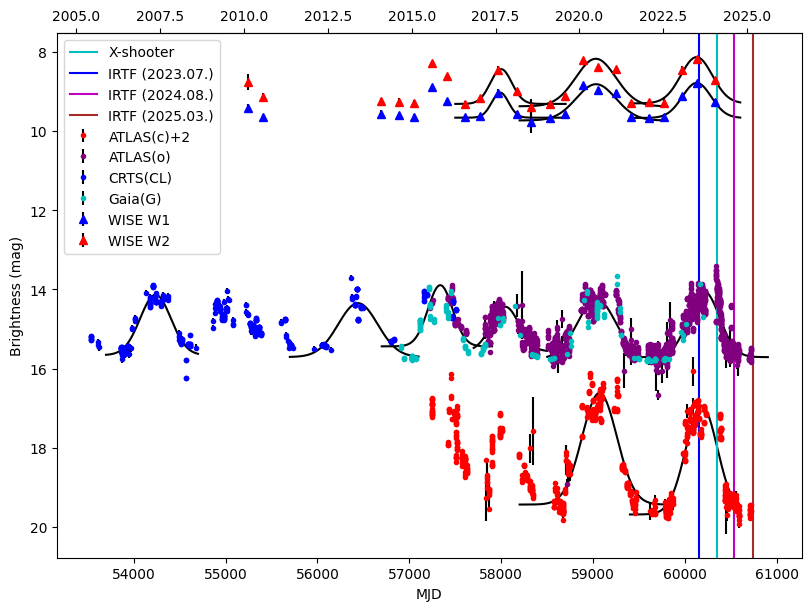}
\includegraphics[width=8.9cm]{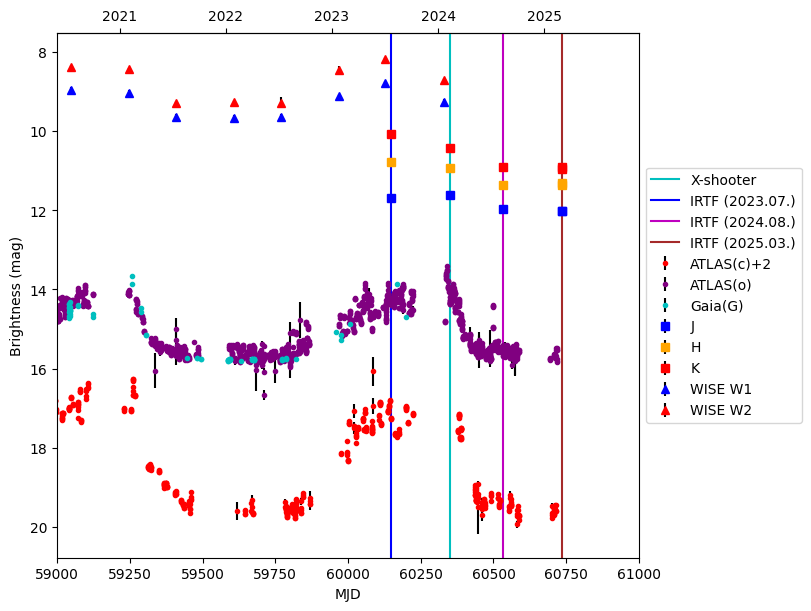}
\caption{\textit{Left panel:} Optical light curve of Gaia21bja based on ATLAS, Gaia $G$, and CRTS data covering about 20 years. \textit{Right panel:} Gaia and ATLAS light curve covering about 5 years with the epochs of the IRTF and X-shooter spectra overplotted. 
}
\label{fig:lightcurve}
\end{figure}

\begin{figure}[h]
    \centering
    \includegraphics[width=0.5\linewidth]{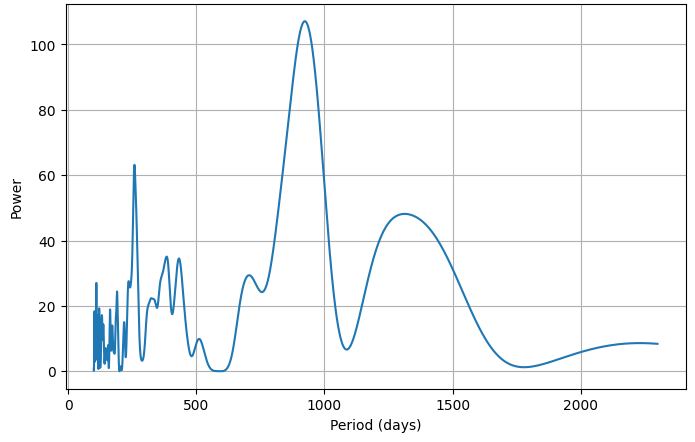}
    \caption{Periodogram based on the CRTS and Gaia $G$ magnitudes.}
    \label{fig:pdg}
\end{figure}

Figure \ref{fig:lightcurve} shows the long-term light curve of Gaia21bja, covering about 20 years, showing at least seven brightenings with amplitudes up to $\sim$1.7 mag in the optical and durations of 1.5-2 years.
By visual inspection, the brightenings may have a quasi-periodic behavior. In order to test this, we performed a Lomb-Scargle periodogram analysis based on the CRTS and Gaia $G$ photometry obtained after 2011, using the public tools at the NASA Exoplanet Archive Periodogram Service\footnote{https://exoplanetarchive.ipac.caltech.edu/cgi-bin/Pgram/nph-pgram}. The result is shown in Fig.~\ref{fig:pdg}. The most significant period was found to be $T=916 \pm 77$ days, where we derived the error of the period by fitting a Gaussian to the peak of the periodogram.
We used the half of the full width at half maximum (FWHM) value of the fitted Gaussian as the uncertainty of the period.

In order to compare the amplitudes and durations of the bursts, we fitted a Gaussian to most peaks. Exceptions were the 2010 peak in the CRTS, the 2015 peak in the NEOWISE, and the 2017 peak in the ATLAS $c$ light curves. The 2010 peak in the CRTS light curve is not well defined, while a fit to the other two peaks was attempted, but the results had large uncertainties.  
The results are summarized in Table \ref{tab:gausssfit}. Based on the Gaia $G$ and CRTS photometry, the largest amplitude corresponds to the 2023 burst and is $\sim$1.7 mag, while the smallest amplitude burst was the one in 2017 with an amplitude of $\sim$1.1 mag. Based on the Gaia $G$ and CRTS data, the 2013 and the 2020 bursts had similar time scales with FWHM of $\sim$215 days and $\sim$210 days, respectively, while the 2015 and 2017 bursts had similar, shorter time scales with FWHM of $\sim$135 days and $\sim$152 days, respectively. Based on the 2017, 2020, and 2023 bursts, the WISE light curves show slightly lower amplitudes compared to the optical light curves. 
The $W2$ light curves show $\sim$30-40\% larger amplitudes compared to the $W1$ light curves.

By visual inspection, the optical and infrared light curves seem to be correlated. To quantify this correlation, we interpolated the Gaia $G$ magnitudes at the epochs of the NEOWISE data, and calculated the Pearson correlation coefficients, which are 0.86 between the interpolated Gaia $G$ and $W1$ magnitudes, and 0.88 between the interpolated Gaia $G$ and $W2$ magnitudes. We discuss the positions of the (NEO)WISE data in Appendix \ref{sec:wise_data}.

\begin{table}[]
\centering
\caption{Gaussian fits to bursts.}
\label{tab:gausssfit}
\begin{tabular}{rllll}
\toprule
Year & Band & Amplitude&      Center&   FWHM\\
     &      &     (mag)& (MJD$-$54000)& (days)\\
\midrule
2007 & CRTS / Gaia G / ATLAS(o) & 1.53 $\pm$ 0.07 & 239 $\pm$ 6 & 168 $\pm$ 12 \\
2013 & CRTS / Gaia G / ATLAS(o) & 1.36 $\pm$ 0.04 & 2441 $\pm$ 12 & 215 $\pm$ 11 \\
2015 & CRTS / Gaia G / ATLAS(o) & 1.55 $\pm$ 0.10 & 3336 $\pm$ 6 & 135 $\pm$ 9 \\
2017 & CRTS / Gaia G / ATLAS(o) & 1.13 $\pm$ 0.05 & 4043 $\pm$ 4 & 152 $\pm$ 8 \\
2017 & WISE W1 & 0.63 $\pm$ 0.12 & 3988 $\pm$ 47 & 92 $\pm$ 29 \\
2017 & WISE W2 & 0.88 $\pm$ 0.15 & 3998 $\pm$ 31 & 117 $\pm$ 26 \\
2020 & CRTS / Gaia G / ATLAS(o) & 1.51 $\pm$ 0.05 & 5039 $\pm$ 4 & 210 $\pm$ 9 \\
2020 & ATLAS(c) & 2.82 $\pm$ 0.07 & 5069 $\pm$ 5 & 189 $\pm$ 6 \\
2020 & WISE W1 & 0.92 $\pm$ 0.13 & 5031 $\pm$ 33 & 217 $\pm$ 42 \\
2020 & WISE W2 & 1.20 $\pm$ 0.18 & 5029 $\pm$ 35 & 218 $\pm$ 45 \\
2023 & ATLAS(c) & 2.75 $\pm$ 0.06 & 6180 $\pm$ 4 & 172 $\pm$ 5 \\
2023 & CRTS / Gaia G / ATLAS(o) & 1.66 $\pm$ 0.02 & 6207 $\pm$ 2 & 168 $\pm$ 3 \\
2023 & WISE W1 & 0.89 $\pm$ 0.04 & 6127 $\pm$ 8 & 157 $\pm$ 9 \\
2023 & WISE W2 & 1.17 $\pm$ 0.12 & 6125 $\pm$ 18 & 171 $\pm$ 23 \\
\bottomrule
\end{tabular}
\end{table}

\begin{figure}
\centering
\includegraphics[width=0.6\linewidth]{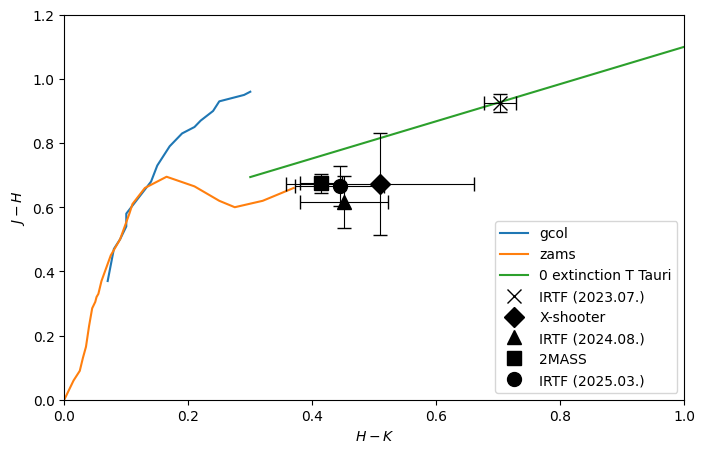}
\caption{$J-H$ versus $H-K$ color-color diagram. The orange and blue lines correspond to the color of the zero-age main sequence and the giant branch \citep{BessellBrett1988}, respectively. The green line shows the locus of unreddened CTTS \citep{Meyer1997}.
}
\label{fig:colorcolor}
\end{figure}

Figure \ref{fig:colorcolor} shows the $J-H$ versus $H-K$ color-color diagram based on the 2MASS magnitudes, the IRTF photometry, and synthetic photometry based on the X-shooter spectrum, which can be used to estimate the visual extinction. All of the data points are close to the locus of unreddened T Tauri stars, suggesting that Gaia21bja has negligible extinction which does not vary significantly. 
This result is consistent with the value derived by \citet{Luhman_2020}, and may suggest that the variability is not due to changes in extinction.

\subsection{Spectral energy distribution}
\label{sec:sed}

\begin{figure}
\centering
\includegraphics[width=0.9\linewidth]{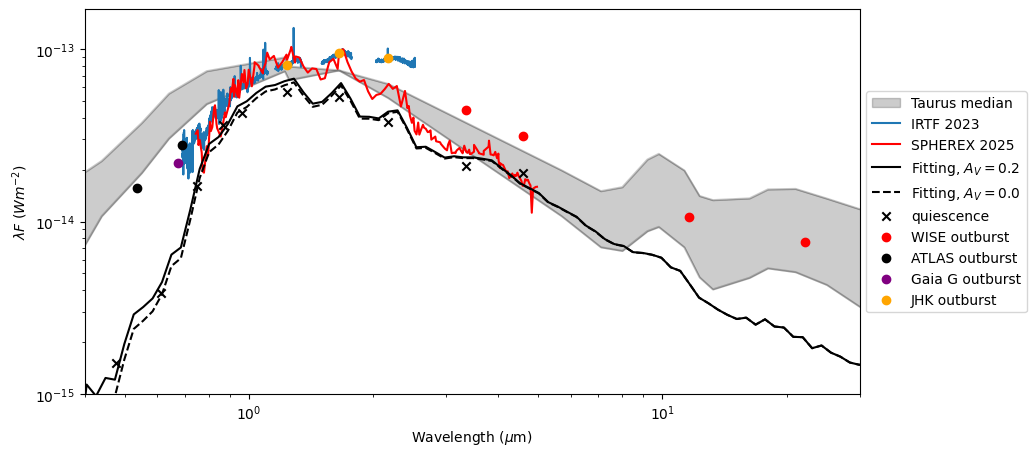}
\caption{SEDs of Gaia21bja in quiescence (crosses) and during the burst (dots). Spectra from IRTF (blue) and SPHEREx (red) are also overplotted. The ‘Taurus median’ SED, normalized to the flux at 1.65 $\mu m$ is overplotted for comparison. The black lines correspond to the best fitting model from \citet{Robitaille2006} corresponding to visual extinctions of 0 mag and 0.2 mag.
}
\label{fig:SED}
\end{figure}

We plotted the Spectral Energy Distribution (SED) of Gaia21bja both in quiescence and during the burst in Figure \ref{fig:SED}. 
To construct the SED in quiescence, we used data from Pan-STARRS PS1 \citep{Magnier2013}, 2MASS, and the faintest (NEO)WISE data points from VizieR. As investigated in Appendix \ref{sec:wise_data}, we consider the (NEO)WISE data points obtained during quiescence as upper limits.
For the SED during the burst, we used data from Gaia $G$ and ATLAS during the 2023 burst, $JHK$ magnitudes obtained with the IRTF in 2023, and the brightest (NEO)WISE data points from VizieR. We overplotted the IRTF spectrum obtained during the burst in 2023, and a spectrum obtained by the Spectro-Photometer for the History of the Universe, Epoch of Reionization and Ices Explorer (SPHEREx, \citealp{Crill2020}, \citealp{irsa629}) in 2025 July-August.
In Figure \ref{fig:SED}, we overplotted the ‘Taurus median’, which is based on the SEDs of CTTS in the Taurus region (\citealp{DAlessio1999}, \citealp{Furlan2006}). The shaded area includes the median as well as the lower and upper quartiles, which define the range around the median where 50\% of the flux values lie \citep{Furlan2006}.

We calculated the spectral index using the formula by \cite{Kuhn_2021} using the $[4.5]-[24]$ color index, as this is the one preferred by the article. We used the photometry from \cite{Luhman_2020} and obtained $\alpha=-0.51$, which corresponds to Class II \citep{Lada1987}.

In order to characterize the quiescent SED, we used the SED Fitter of \citet{Robitaille2007} and the model grid of \citet{Robitaille2006}. This model grid of YSOs is based on the radiation transfer code by \citep{Whitney2003a,Whitney2003b}. The 200,000 SED models cover a large range of stellar masses (from 0.1 to 50 $M_\odot$) and evolutionary stages (from embedded protostars to transitional disks). The model consists of a pre-main sequence star surrounded by an accretion disk and a
rotationally flattened envelope with bipolar cavities. Scattering and reprocessing of the stellar
radiation by dust, as well as the accretion luminosity are taken into account in the model.
The SED Fitter fits each of the models to the data, allowing the $A_V$ to be a free parameter, and performs a $\chi^2$ minimization. The best fit model, overplotted in Figure \ref{fig:SED} was found to have an $A_V$ of 0.2 mag, however, an $A_V$ of 0 mag provides a similar fit. The best fit effective temperature of 2982~K, found for the model fit, is close to the $3190\pm75$ K derived by \citet{Thanathibodee2022}.
The very low $A_V$ found in this model fit is consistent with the result from the $J-H$ versus $H-K$ color-color diagram. 
Assuming an $A_V$ of 0.2 mag has very little effect on the results in Sections \ref{sect:stellar_param} and \ref{sect:accr_param}. Furthermore, we investigated, that the mass accretion rate, which is estimated in Sect. \ref{sect:accr_param}, is most consistent with an $A_V$ of 0 mag. Therefore, for the calculations in the next sections, we assume an $A_V$ of 0 mag.

\subsection{Line detections}
\label{sec_line_det}

\begin{figure*}[htbp]
\centering
\includegraphics[width=0.9\textwidth]{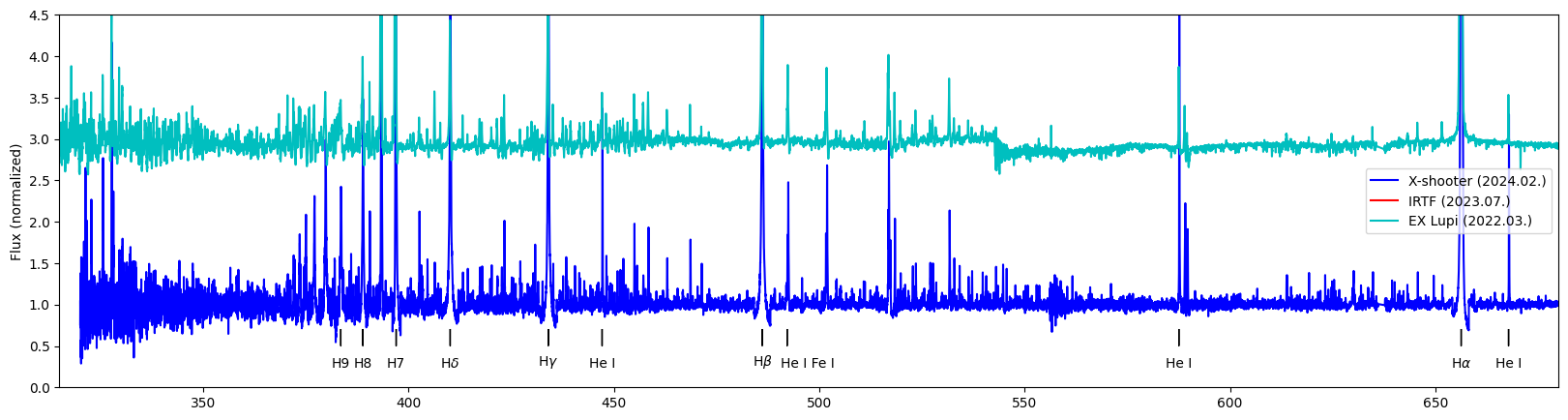}
\includegraphics[width=0.9\textwidth]{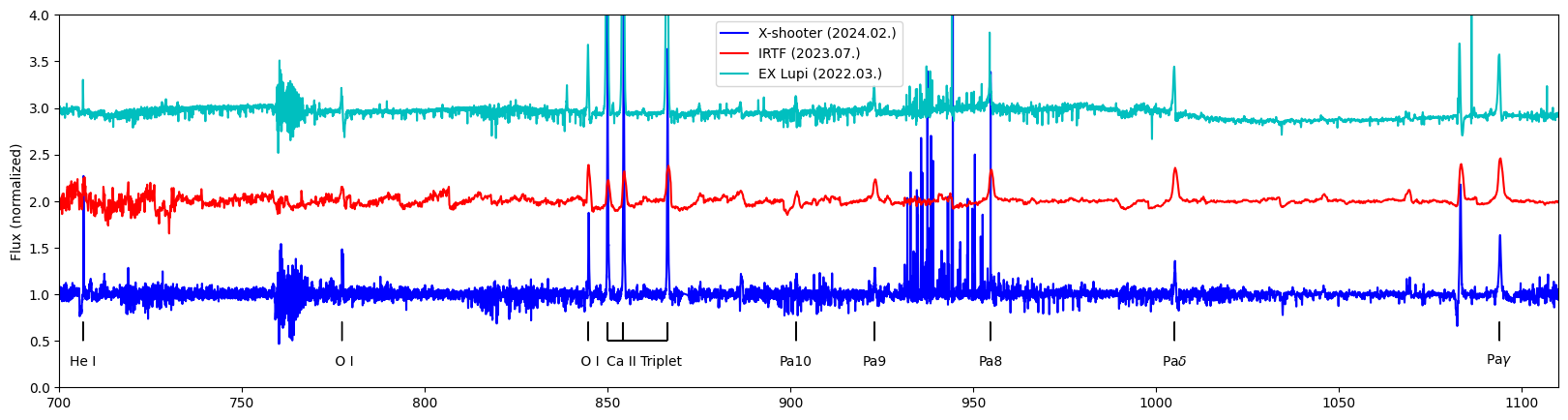}
\includegraphics[width=0.9\textwidth]{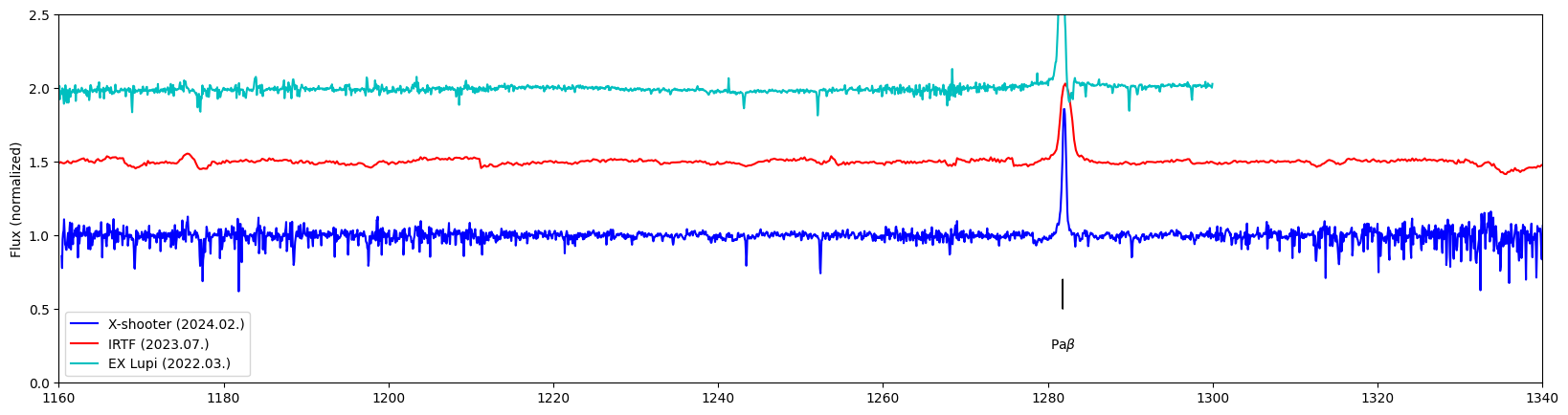}
\includegraphics[width=0.9\textwidth]{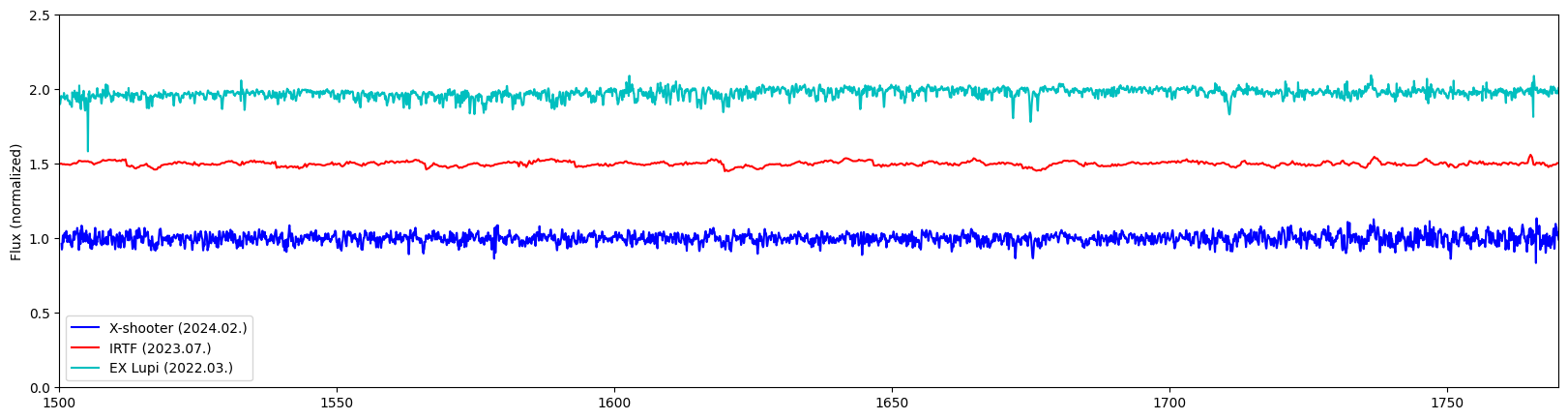}
\includegraphics[width=0.9\textwidth]{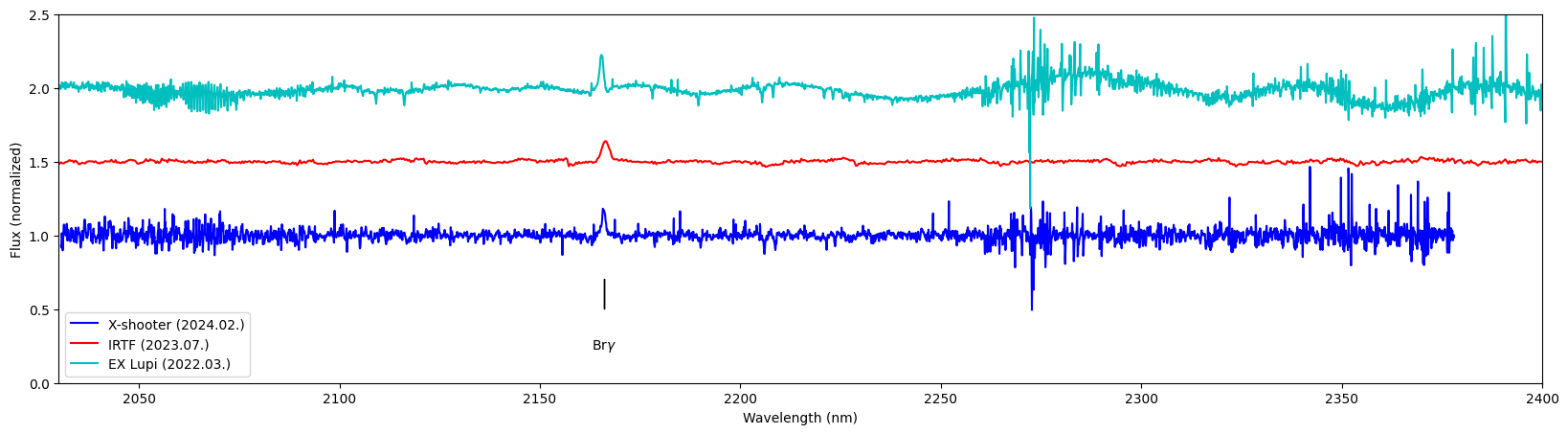}
\caption{Spectra of Gaia21bja observed during the burst using the IRTF/SpeX and the VLT/X-shooter. We also plotted the outburst spectrum of EX Lupi as a reference.
}
\label{fig:outburst_spectra}
\end{figure*}

\begin{figure*}[htbp]
\centering
\includegraphics[width=\textwidth]{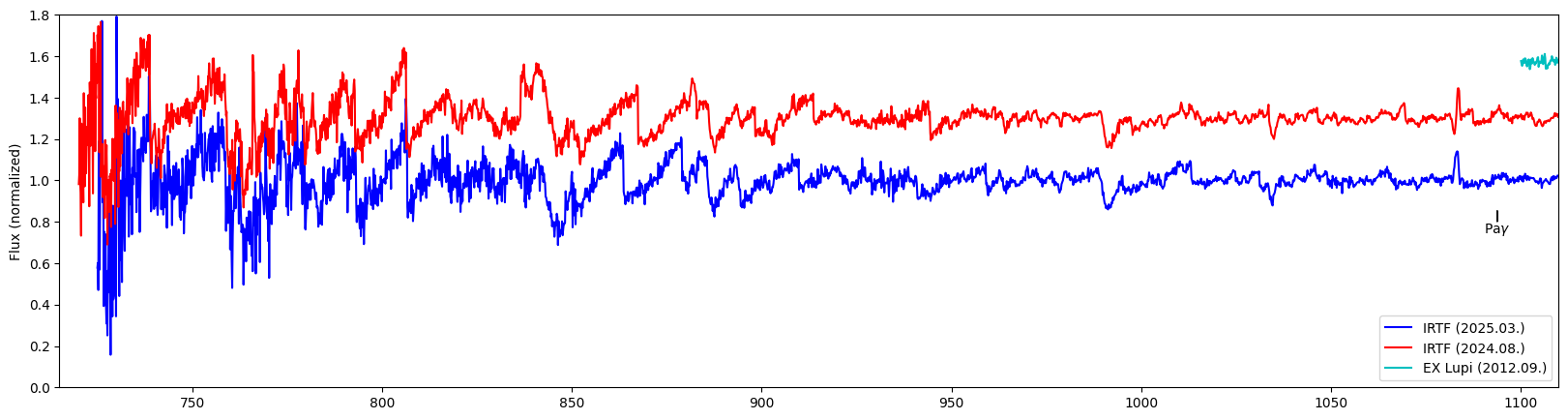}
\includegraphics[width=\textwidth]{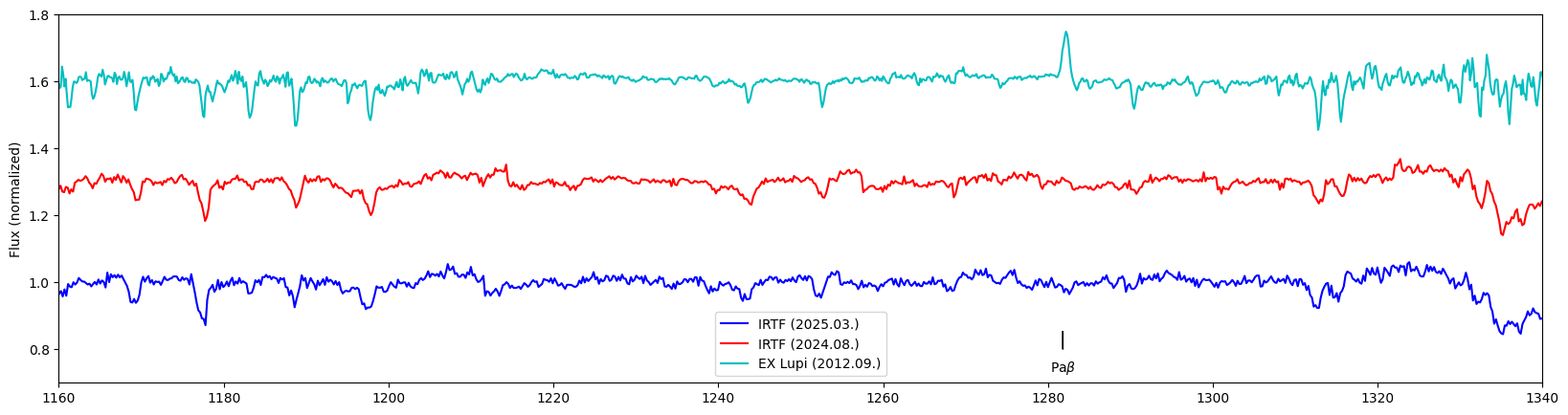}
\includegraphics[width=\textwidth]{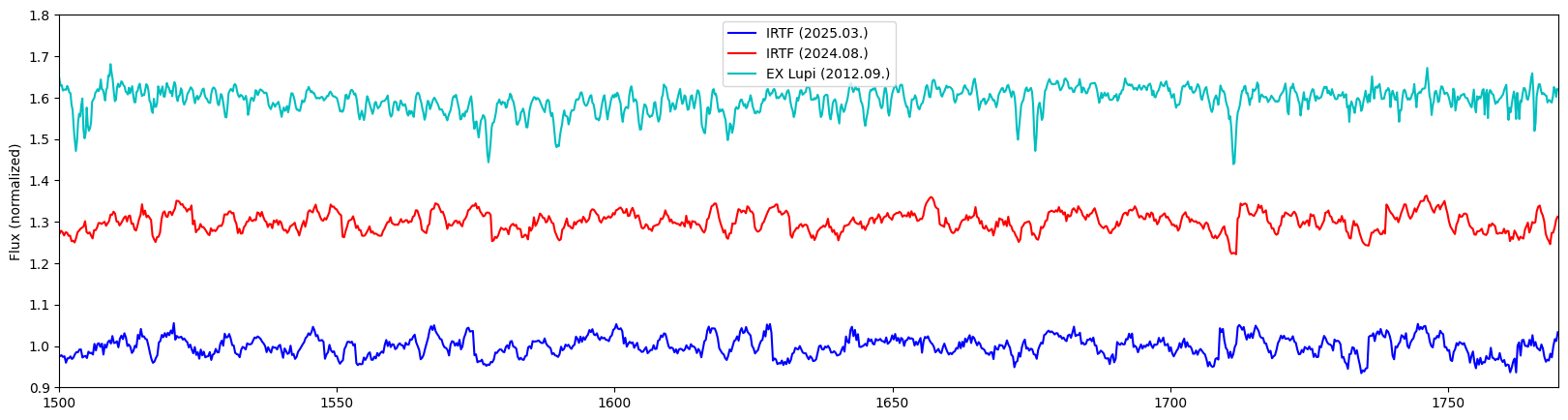}
\includegraphics[width=\textwidth]{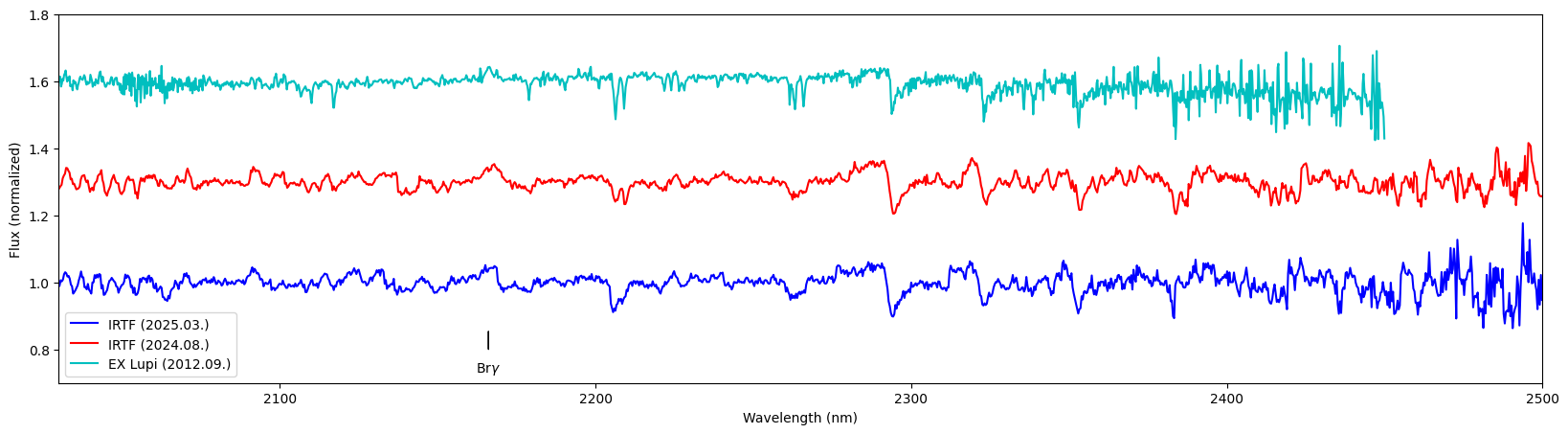}
\caption{Spectra of Gaia21bja observed during quiescence using IRTF/SpeX. We also plotted the quiescence spectrum of EX Lupi as a reference.
}
\label{fig:quiescence_spectra}
\end{figure*}

The spectra taken during quiescence with the IRTF/Spex and during burst with the IRTF/Spex and VLT/X-shooter are shown in Figures \ref{fig:outburst_spectra} and \ref{fig:quiescence_spectra}. 
In both the X-shooter and IRTF spectrum we discarded values in the wavelength ranges 1.11 $\mu$m -- 1.16 $\mu$m, 1.34 -- 1.50 $\mu$m, and 1.77 -- 2.03 $\mu$m, as these are heavily affected by atmospheric effects.
We also applied a correction from vacuum to air in the case of the outburst IRTF spectrum using the formula by \cite{2000ApJS..130..403M}, as the other two spectra are given in air.
For a comparison, we also plotted the X-shooter spectra of EX Lupi during its recent burst \citep{CruzSaenzdeMiera2023} and quiescence \citep{Rigliaco2020}.
The IRTF and VLT spectra taken during the burst are similar to those of EXors and CTTS. The CO-bandhead is seen in absorption in quiescence and during the burst. There are various accretion rate tracer lines detected at both epochs during the burst, including the Balmer, Brackett and Paschen series of H\,I, Ca\,II, O\,I, He\,I, and Na\,I. 
However, only a very few and faint accretion tracers are seen in quiescence, suggesting a low mass accretion rate.
The line fluxes during the burst and in quiescence are listed in Tables \ref{tab:fluxes_accrates} and \ref{tab:fluxes_accrates_q}, respectively.

\subsection{Stellar parameters}
\label{sect:stellar_param}

\begin{figure}
\centering
\includegraphics[width=0.5\linewidth]{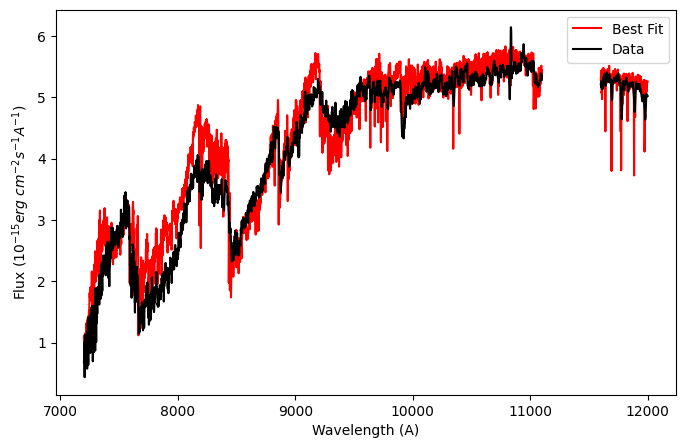}
\caption{The IRTF spectrum obtained in August 2024 (during quiescence) and the best fitting BT-Settl model, corresponding to a temperature of $T = 3000$ K.}
\label{fig:T_fit}
\end{figure}

We use the quiescence spectra obtained with the IRTF in August 2024 to estimate the effective temperature of the star by comparing it to BT-Settl (AGSS2009) photospheric templates (\citealp{2011ASPC..448...91A}, \citealp{2012RSPTA.370.2765A}, \citealp{2009ARA&A..47..481A}, \citealp{2006MNRAS.368.1087B}). We removed parts of the dataset affected by the telluric correction (as indicated in Section \ref{sec_line_det}), as well as data above 1200 nm, as these are affected by the accretion disk.
The photospheric templates provide the flux emitted by a star at a given temperature, $\log(g)$, metallicity, and wavelength, which we interpolated whenever needed. We compared the models with $\log(g)=4$ and $\mathrm{[Fe/H]}=0.37$ assuming $A_V=0$ mag to the observed quiescence spectra. 
We used non-linear least-squares minimization (LMFIT library) in Python for fitting, and the best fitting model corresponds to a temperature of $T = 3000$ K as seen in Figure \ref{fig:T_fit}. As we had grid points every 100 K, we estimate the error to be 50 K. This result is also close to the previous SED fitting and the earlier reported value of $3190\pm75$~K \citep{Thanathibodee2022}.

Once the effective temperature and visual extinction are estimated, the stellar luminosity, mass, and radius can be derived following a method previously used by \citet{Fiorellino2021}. The stellar luminosity can be derived as
\begin{equation}
    L_\star = 10^{0.4(M_{\rm{bol},\odot} - M_{\rm{bol}})} L_\odot
    \label{eq:L*}
\end{equation}
$M_{\rm{bol},\odot}$ is the bolometric magnitude of the Sun \citep{Mamajek2015}, and the absolute bolometric magnitude of the star, $M_{\rm{bol}}$ can be calculated as 
\begin{equation}
    M_{\rm{bol}} = (J+BC_J) + 5 - 5\log(d[\rm{pc}])
    \label{eq:Mbol}
\end{equation}
where $J$ is the extinction corrected $J$ band magnitude in quiescence (in general, however, for Gaia21bja, no extinction correction was needed) and $BC_J$ is the $J$ band bolometric correction. We used the $BC_J$ estimated for 5-30 Myr stars by \citet{PecautMamajek2013} for M3 type stars (based on the effective temperature derived above). Based on the above, we calculated $L_\star=(4.5\pm0.3) \times 10^{-2}~L_\odot$.

To investigate the effect of using an optical band to estimate the stellar luminosity, we used the lowest $V$-band magnitude found in archival data: $V=16.4$ mag \citep{Zacharias2013} and the $V$-band bolometric correction estimated for 5-30 Myr stars by \citet{PecautMamajek2013}. 
Assuming a 10\% error for the $V$ magnitude, we obtained $L_\star=(6.4\pm1.3) \times 10^{-2}~L_\odot$. Since this result is fairly consistent with the one based on the $J$ magnitude, we continue to use the values obtained from the $J$ magnitude in the following.

Based on the $L_\star=(4.5\pm0.3) \times 10^{-2}~L_\odot$, the effective temperature derived above, and the evolutionary tracks by \citet{Siess2000}, we determined a stellar mass of $M_\star= 0.16 \pm 0.03~M_\odot$, which corresponds to a metallicity of Z=0.04 and an age of $5.97\times10^6$ yr.
The stellar radius can be derived as
\begin{equation}
\label{eq:R*}
R_\star = \frac{1}{2T^2}\sqrt{\frac{L_\star}{\pi\sigma}}
\end{equation}
where $T$ is the effective temperature and $\sigma$ is the Stefan-Boltzmann constant. This results in a stellar radius of $R_\star= 0.78 \pm 0.04~R_\odot$. In the following, we can use the stellar parameters derived in this section to estimate the accretion parameters.

\subsection{Accretion parameters}
\label{sect:accr_param}

The fluxes of the accretion tracer lines listed in Table \ref{tab:fluxes_accrates} can be used to estimate the accretion luminosities. Since we found the extinction to be negligible, these fluxes ($f_{\rm{line}}$) can be directly converted to line luminosities as
\begin{equation}
    L_{\rm{line}}=4 \pi d^2 f_{\rm{line}}
    \label{eq:Lline}
\end{equation}
which can be converted to accretion luminosities $L_{\rm{acc}}$ using the empirical relations from \cite{Alcala2017}. Assuming an inner-disk radius of 5 $R_\star$ \citep{Hartmann1998}, the mass accretion rate can be calculated as
\begin{equation}
   \dot{M}_{\rm{acc}} = 1.25\frac{L_{\rm{acc}}R_\star}{GM_\star}.
   \label{eq:accretion}
\end{equation}
where $R_\star$ and $M_\star$ are the radius and mass of the star, respectively.

\begin{figure}[htbp]
\centering
\includegraphics[width=0.6\linewidth]{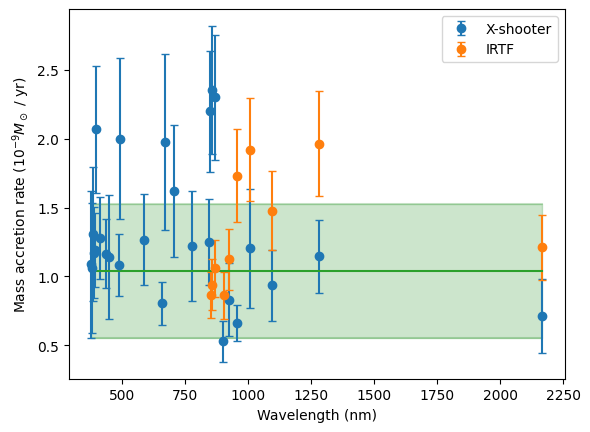}
\caption{Mass accretion rates during the burst obtained from the line luminosities and the empirical relations of \citet{Alcala2017}. The green line is the average of all the data points weighted by variance. The shaded green area shows the values which are within the standard deviation.}
\label{fig:acc_rate}
\end{figure}

Since we detected some emission lines in quiescence, we could calculate the mass accretion rate both in quiescence and during the burst.
The values during the burst are plotted in Figure \ref{fig:acc_rate} and are listed in Table \ref{tab:fluxes_accrates}. We have discarded lines where the fluxes have more than a 50\% error (SNR<2). We did not take into account the He I line at $\lambda=501.57$ nm, as its flux may trace winds/outflows \citep{Edwards_2006}.
We also considered visual extinctions above the assumed 0 mag, up to an $A_V$ of 0.5 mag, however, we found the mass accretion rates estimated from the different lines are more consistent with no extinction rather than values of up to an $A_V$ of 0.5 mag, confirming our assumption based on the $J-H$ vs $H-K$ diagram and the SED fitting, that extinction toward Gaia21bja is indeed negligible.

We consider the accretion luminosity and mass accretion rate at each epoch as the variance-weighted average of the values derived from the different lines and their standard deviation as the error. 
In quiescence, the accretion luminosity and the mass accretion rate were $(3.7\pm 5.5)\times^{-4}L_\odot$ and $(5.3\pm10.7)\times10^{-11}$ $M_\odot$ yr$^{-1}$ in 2024 August and $(1.7\pm 0.1)\times10^{-3}L_\odot$ and $(3.4\pm0.7)\times10^{-10}$ $M_\odot$ yr$^{-1}$ in 2025 March, respectively. 
During the burst, we obtained an accretion luminosity of $(4.6\pm2.6)\times10^{-3}$ $L_\odot$ and a mass accretion rate of $(1.0\pm0.5)\times10^{-9}$ $M_\odot$ yr$^{-1}$ at the epoch of the X-shooter spectrum, and an accretion luminosity of $(6.4\pm2.2)\times10^{-3}$ $L_\odot$ and a mass accretion rate of $(1.1\pm0.4)\times10^{-9}$ $M_\odot$ yr$^{-1}$ at the epoch of the IRTF spectrum, suggesting that the accretion luminosity and the mass accretion rate did not change during the burst.
Taking the average of the accretion luminosities derived from the quiescent spectra, $\sim1.0\times10^{-3}$ $L_\odot$, and the average of the accretion luminosities derived during the burst, $\sim5.5\times10^{-3}$ $L_\odot$, the accretion luminosities increased by a factor of $\sim$5.5 between the quiescent and burst states.

To obtain another estimate of the accretion parameters, we used the SED Fitter of \citet{Robitaille2007}, as in Sect. \ref{sec:sed} for the quiescent SED, and fit also the SED during the burst, in order to obtain bolometric luminosities. The bolometric luminosity given by the model is $9.9\times10^{-2}~L_\odot$ during quiescence and $3.3\times10^{-1}~L_\odot$ during the burst.
Assuming $L_{\rm{acc}} \sim L_{\rm{bol}}-L_\star$, the accretion luminosity in quiescence is $\sim0.05~L_\odot$ and it is $\sim0.29~L_\odot$ during the burst, assuming the stellar luminosity of $4.5\times10^{-2}~L_\odot$ obtained above. Using eqn. \ref{eq:accretion}, the corresponding mass accretion rates are $1.1\times10^{-8}$ $M_\odot$ yr$^{-1}$ during quiescence and $5.7\times10^{-8}$ $M_\odot$ yr$^{-1}$ during the burst, which are more than an order of magnitude above the values obtained using the line luminosities. 
The difference between the accretion luminosities derived from the SEDs for quiescence and the burst suggests a factor of $\sim$6 increase between quiescence and the burst, which is similar to what was found for the 2022 burst of EX Lupi \citep{CruzSaenzdeMiera2023}.
We discuss the accretion parameters in Sect. \ref{sect:discussion_acc}.

\section{Discussion}
\label{sect_discussion}

We analyzed the optical and near-infrared photometry and spectroscopy of the young stellar object Gaia21bja. There are several brightening events with an amplitude up to $\sim$1.7 mag in Gaia $G$-band. The duration of these brightening events is about 1.5--2 years. Based on the time scales and amplitudes, the brightenings of Gaia21bja may be consistent with eruptive YSO bursts. In addition to the light curve, its spectra during the burst are similar to those of EXors. Here we discuss the accretion luminosities and mass accretion rates, and the quasi-periodic variability.

\subsection{Accretion parameters}
\label{sect:discussion_acc}

In Sect. \ref{sect:accr_param} we derived accretion parameters using two different methods and found that the accretion parameters increased by a factor of $5.5-6$ during the burst. 
Considering the line luminosities of the detected emission lines and converting them to accretion luminosities using the \citet{Alcala2017} empirical relations resulted in more than an order of magnitude lower accretion luminosities compared to those obtained from bolometric luminosities derived by SED fitting. 
The \citet{Alcala2017} empirical relations assume that the magnetospheric accretion model is applicable, which is the case for CTTS \citep{Hartmann2016}. Magnetospheric accretion is also expected in the case EXor bursts and outbursts \citep{SiciliaAguilar2015}. However, recent studies questioned the validity of magnetospheric accretion for (out)bursts (\citealp{CruzSaenzdeMiera2023,Fiorellino2025}). In particular, \citet{CruzSaenzdeMiera2023} found that accretion luminosities derived for EX Lupi using the empirical relations and those obtained from a slab model may be up to an order of magnitude different. 
The inconsistency between the results of the two methods used to estimate the accretion luminosities suggests, that similarly to the cases studied by \citet{CruzSaenzdeMiera2023} and \citet{Fiorellino2025}, the magnetospheric accretion model may not be applicable to the burst of Gaia21bja.

\begin{figure}[htbp]
\centering
\includegraphics[width=0.7\linewidth]{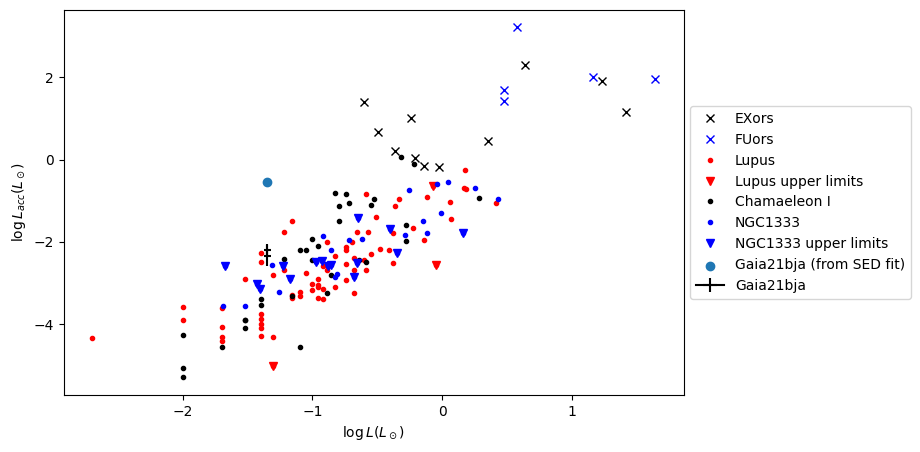}
\includegraphics[width=0.7\linewidth]{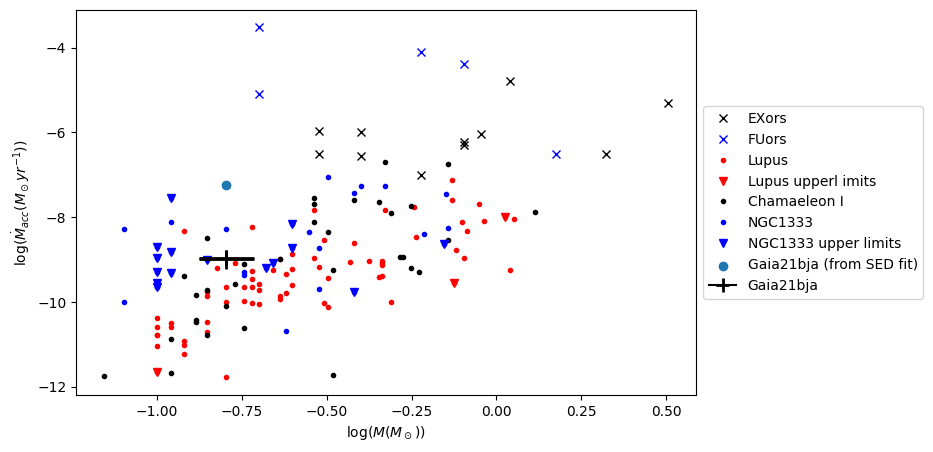}
\caption{Accretion luminosities ($L_\odot$) versus stellar luminosities ($L_\odot$) (top panel) and mass accretion rates ($M_\odot$ yr$^{-1}$) versus stellar masses ($M_\odot$) (bottom panel) and their comparison to confirmed EXors and FUors (crosses) and samples of CTTS in the Lupus (red symbols, \citealp{Alcala2019}), Chamaeleon I (black symbols, \citealp{Manara2019}), and NGC 1333 (blue symbols, \citealp{Fiorellino2021}) regions. The upper limits are marked with downward triangles. The parameters for the EXors and FUors obtained during their outbursts are based on the OYCAT \citep{ContrerasPena2025} and references therein.}
\label{fig:acc_stellar_param}
\end{figure}

In Figure \ref{fig:acc_stellar_param} we compare the accretion luminosities and mass accretion rates to the stellar luminosity and mass, respectively, for Gaia21bja using both methods and compare them to those derived for samples of CTTS and for confirmed eruptive YSOs.
For CTTS we used samples towards the Lupus (black symbols, \citealp{Alcala2019}), the Chamaeleon I (grey symbols, \citealp{Manara2019}), and the NGC 1333 (light blue symbols, \citealp{Fiorellino2021}) regions. The parameters for the EXors and FUors are based on the Outbursting YSOs Catalogue (OYCAT, \citealp{ContrerasPena2025}) and references therein.
The accretion parameters derived from the bolometric luminosity are clearly above the CTTS with similar luminosities and masses. Even the accretion parameters estimated using the empirical relations are among the highest values for similar stellar parameters.

\begin{figure}
    \centering
    \includegraphics[width=0.5\linewidth]{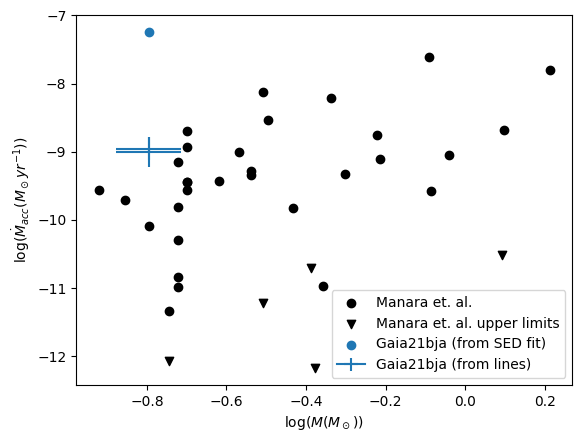}
    \caption{Mass accretion rates versus stellar masses in the Upper Scorpius \citep{Manara2020}.}
    \label{fig:upperSco}
\end{figure}

We also compare the derived accretion parameters to those corresponding to the same star-forming region.
As a member of Upper Sco, Gaia21bja is most likely an evolved, 5-10 Myr old, pre-main sequence star. This is supported by Fig. 2 of \citet{Manara2020}, as the luminosity and effective temperature of Gaia21bja are consistent with the HR diagram of Upper Sco (Gaia21bja is located at log~$T_{\rm{eff}} (K) = 3.477$, log~$L_\star (L_\odot)=-1.35$). Based on the H$\alpha$ line, observed on 2017 July 17, \citet{Thanathibodee2022} selected this star into a list of low accretors.
In Figure \ref{fig:upperSco} we plotted the mass accretion rate versus stellar mass of Gaia21bja derived using both methods together with those derived for the members of Upper Sco by \citet{Manara2020} and shown in their Table 1. 
Figure \ref{fig:upperSco} shows that Gaia21bja is among the lowest mass members of the cluster among those with mass accretion rate estimates. The mass accretion rate of Gaia21bja derived using both methods during its burst is among the highest values for a similar mass in the Upper Sco region.
Based on the above comparisons, we suggest that Gaia21bja is an eruptive YSO.
Based on its quasi-periodic light curve, Gaia21bja seems to be most consistent with the `Periodic' group of the OYCAT \citep{ContrerasPena2025}.

\subsection{Quasi-periodic variability}

An interesting characteristics of this star is the quasi-periodic occurrence of the brightening events, which was found for 8 eruptive YSOs so far based on the new catalog by \citet{ContrerasPena2025}. The cycle lengths of these brightening events cover a large range from $\sim$25 days for LRLL54361 \citep{Muzerolle2013,LeGouellec2024} to 1437 days for VVVv32 \citep{ContrerasPena2017,Guo2022}. The period derived above for Gaia21bja falls in this range: $916\pm77$ days. It was suggested, that these quasi-periodic changes may be due to a stellar or planetary-mass companion interacting with the disk \citep{Hodapp2012,Yoo2017,Guo2022}. 
Assuming this is the case, we can estimate the semi-major axis of the orbit using Kepler's 3rd law
\begin{equation}
    \frac{a^3}{T^2}=\frac{GM}{4\pi^2},
\end{equation}
where $T$ is the period and $M$ is the total mass. Assuming that the companion's mass $m$ is much smaller than that of the main star ($M_*$) we get $a=1.00\pm0.08~AU$. This gives a lower limit for the semi-major axis. However, if $m$ would be comparable to $M_*$, then the wiggling of the main component would be of the order of $a$ and be detectable by Gaia. Since Gaia has found no signs of a companion, we accept that $m\ll M_*$.

We can also use the fact that Gaia could not detect a component to give an upper limit for the companion's mass. According to \citet{MignardKlioner2010}, the astrometric accuracy of Gaia is $\theta_{\rm{lim}} \sim 130~\mu as$ for sources with Gaia magnitude of 16. Thus, the smallest amount of wiggling Gaia can detect is $\Delta x_{\rm{lim}}=\theta_{\rm{lim}}d$, where $\theta_{\rm{lim}}$ is in radians.\\
The semi-major axis of the main component's orbit is
\begin{equation}
    a_* = \frac{m}{M_*+m}a\approx\frac{m}{M_*}a.
\end{equation}
In the worst case the orbit is oriented such that the minor axis is perpendicular to the line of sight and the main component's wiggle is $\Delta x=2b_*$, where $b_*=\sqrt{1-e^2}~a_*$ is the semi-minor axis of the main component's orbit, where $e$ is the eccentricity. Putting these together, we get
\begin{align*}
    \Delta x_{\rm{lim}}&\geq 2b_*\\
    \theta_{\rm{lim}}d&\geq 2\sqrt{1-e^2}\frac{m}{M_*}a\\
    m\sqrt{1-e^2}&\leq M_*\frac{\theta_{\rm{lim}}d}{2a}=1.6~M_{\rm{Jupiter}}
\end{align*}
where $M_{\rm{Jupiter}}$ is the mass of Jupiter. Typically $0.1<\sqrt{1-e^2}< 1$, which is equivalent with an upper limit of around $1-10~M_{\rm{Jupiter}}$. 
This suggests that the component (if exists) is a planetary component.

\section{Summary}
\label{sect_summary}

We studied the photometry and spectroscopy of a Gaia alerted YSO, Gaia21bja. Our main results are summarized below.

\begin{itemize}

\item[-] The 20-year-long optical light curve of the star shows at least seven bursts with an amplitude up to 1.7 mag, which appear to be quasi-periodic. Using a Lomb-Scargle periodogram analysis, we derived the most significant period to be $916\pm77$ days.

\item[-] Based on a comparison to photospheric templates, we estimated the effective temperature of the star to be $T = 3000 \pm 50$ K. We derived the stellar radius, luminosity, and mass to be $R_\star= 0.78 \pm 0.04~R_\odot$, $L_\star=(4.5\pm0.3) \times 10^{-2}~L_\odot$, and $M_\star= 0.16 \pm 0.03~M_\odot$.

\item[-] We estimated accretion luminosities and mass accretion rates using two different methods, and found that the accretion parameters of the star increased by a factor of $5.5-6$ during the burst.

\end{itemize}

Based on the above, we conclude that Gaia21bja is an eruptive YSO, which is most consistent with the `Periodic' category of the OYCAT \citep{ContrerasPena2025}, the first object in this group which was found from the Gaia Alerts.

\vspace{5 mm}
We thank the referee for comments and suggestions which helped to improve this paper.
We acknowledge the Hungarian National Research, Development and Innovation Office grant OTKA FK 146023.
We acknowledge support from the ESA PRODEX contract nr. 4000132054.
G. M. and Zs. N. were supported by the J\'anos Bolyai Research Scholarship of the Hungarian Academy of Sciences.

F. C. S. M. received financial support from the European Research Council (ERC) under the European Union’s Horizon 2020 research and innovation programme (ERC Starting Grant "Chemtrip", grant agreement No 949278). 

The IRTF/SpeX data presented here were obtained at the Infrared Telescope Facility, which is operated by the University of Hawaii under contract 80HQTR24DA010 with the National Aeronautics and Space Administration.

This work was also supported by the NKFIH NKKP grant ADVANCED 149943 and the NKFIH excellence grant TKP2021-NKTA-64. Project no. 149943 has been implemented with the support provided by the Ministry of Culture and Innovation of Hungary from the National Research, Development and Innovation Fund, financed under the NKKP ADVANCED funding scheme.

Based on observations made with ESO Telescopes at the La Silla Paranal Observatory under programme ID 0112.C-0201(A)
or 112.25XA.001.

This project has received funding from the European Research Council (ERC) via the ERC Synergy Grant ECOGAL (grant 855130). Views and opinions expressed are however those of the author(s) only and do not necessarily reflect those of the European Union or the European Research Council Executive Agency. Neither the European Union nor the granting authority can be held responsible for them.

For this work we have used Python in the Google Colaboratory environment and the libraries NumPy (\cite{harris2020array}), SciPy (\cite{2020SciPy-NMeth}, Matplotlib (\cite{Hunter:2007}), Pandas (\cite{reback2020pandas}, \cite{mckinney-proc-scipy-2010}), LMFIT (\cite{newville_2025_16175987}, and Astropy (\cite{astropy:2013}, \cite{astropy:2018}, \cite{astropy:2022}).

We acknowledge ESA \textit{Gaia}, DPAC and the Photometric Science Alerts Team (\url{http://gsaweb.ast.cam.ac.uk/alerts}).

\facilities{VLT, IRTF}

\bibliography{gaia21bja}{}
\bibliographystyle{aasjournal}

%
%
%
%
%
%
%

\appendix

\section{Positions of the (NEO)WISE data}
\label{sec:wise_data}

\begin{figure}[h!]
\centering
\includegraphics[width=0.7\textwidth]{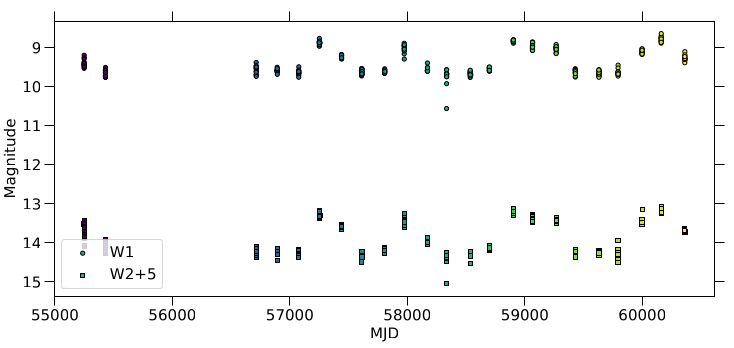}
\includegraphics[width=0.7\textwidth]{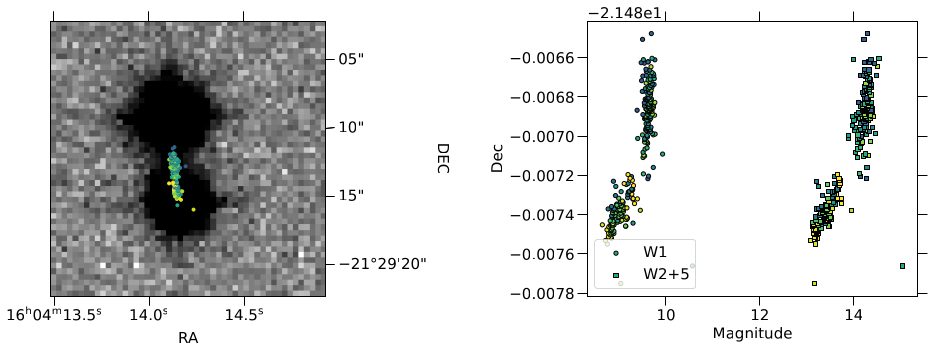}
\caption{\textit{Top panel:} Single epoch (NEO)WISE $W1$ and $W2$ light curves.
\textit{Bottom left panel:} The positions of the (NEO)WISE data points overplotted on a UKIRT Infrared Deep Sky Survey (UKIDSS, \citealp{Lawrence2007}) $J$-band image. North is up and east is to the right. Gaia21bja is the southern source. \textit{Bottom right panel:} (NEO)WISE magnitudes versus declination.}
\label{fig:lightcurve_wise}
\end{figure}

Despite the correlation between the optical and (NEO)WISE light curves (Fig. \ref{fig:lightcurve}), we investigated the positions of the (NEO)WISE data points due to another star (2MASS J16041412-2129089) seen close to Gaia21bja.
Figure \ref{fig:lightcurve_wise} shows single epoch (NEO)WISE $W1$ and $W2$ light curves as well as the positions of the data points.
In the bottom left panel of Figure \ref{fig:lightcurve_wise} it is seen that some of the (NEO)WISE data points were obtained toward Gaia21bja (the bottom source in the figure), however, some of the observations were pointed between Gaia21bja and the star seen nearby. As seen in the bottom right panel of Figure \ref{fig:lightcurve_wise}, which shows the magnitudes versus declinations of the (NEO)WISE data points, in the bright state, the (NEO)WISE source positions are centered on Gaia21bja, while in the faint state, they are between Gaia21bja and 2MASS J16041412-2129089. The reason for this is probably that during the bright state, Gaia21bja is much brighter than 2MASS J16041412-2129089, as indicated by the measured source position close to the nominal position of Gaia21bja. 
During the faint state, the two sources probably have comparable brightness in the W1 and W2 bands, as indicated by the shift in source position, or the 2MASS source may actually be a bit brighter than Gaia21bja. The lower right panel of Fig. \ref{fig:lightcurve_wise}, shows how the declinations of the NEOWISE data points change as a function of their brightness. At the brightest state, the data points are centered on Gaia21bja, while in the faintest state, the offset between the data points from Gaia21bja is up to about 3.5$''$. This is about half of the separation between the two objects, which is about 6.34 arcsec based on the UKIDSS $J$-band image.
Therefore, we consider the \textit{(NEO)WISE} data points obtained during quiescence as upper limits.

\section{Line fluxes}

We present the list of lines which were used to estimate the accretion parameters. 
Tables \ref{tab:fluxes_accrates} and \ref{tab:fluxes_accrates_q} include the line fluxes, accretion luminosities, and mass accretion rates for each accretion tracer obtained during burst and quiescence, respectively.
The line fluxes were calculated by numerically integrating the line profiles using Simpson's rule, and the errors were calculated by propagating the error of the individual data points.
The lines included in Table \ref{tab:fluxes_accrates_q} have a signal to noise ratio of at least 5.

\begin{table*}[h!]
\centering
\caption{Summary of the line fluxes, accretion luminosities, and mass accretion rates during outburst.}
\label{tab:fluxes_accrates}
\begin{tabular}{llrrlll}
\hline
Line& Telescope& $\lambda_{\rm{lab}}$ (nm)& $\lambda$ (nm)& $F$ (W/m$^2$)&	$L_{\rm{acc}}$ ($L_\odot$)& $\dot{M}$ ($M_\odot$/yr)\\
\hline
H$\alpha$ & X-shooter & 656.28 & 656.20 & $(3.35 \pm 0.05) \times 10^{-16}$ & $(4.15 \pm 0.09) \times 10^{-3}$ & $(8.05 \pm 1.57) \times 10^{-10}$ \\
H$\beta$ & X-shooter & 486.13 & 486.06 & $(8.40 \pm 0.53) \times 10^{-17}$ & $(5.58 \pm 0.40) \times 10^{-3}$ & $(1.08 \pm 0.22) \times 10^{-9}$ \\
H$\gamma$ & X-shooter & 434.05 & 433.96 & $(5.61 \pm 0.49) \times 10^{-17}$ & $(6.01 \pm 0.58) \times 10^{-3}$ & $(1.17 \pm 0.25) \times 10^{-9}$ \\
H$\delta$ & X-shooter & 410.17 & 410.10 & $(4.65 \pm 0.54) \times 10^{-17}$ & $(6.59 \pm 0.82) \times 10^{-3}$ & $(1.28 \pm 0.30) \times 10^{-9}$ \\
H7 & X-shooter & 397.01 & 396.80 & $(5.96 \pm 0.61) \times 10^{-17}$ & $(1.07 \pm 0.12) \times 10^{-2}$ & $(2.07 \pm 0.46) \times 10^{-9}$ \\
H8 & X-shooter & 388.90 & 388.84 & $(3.20 \pm 0.60) \times 10^{-17}$ & $(6.05 \pm 1.21) \times 10^{-3}$ & $(1.17 \pm 0.33) \times 10^{-9}$ \\
H9 & X-shooter & 383.54 & 383.48 & $(2.58 \pm 0.79) \times 10^{-17}$ & $(6.74 \pm 2.16) \times 10^{-3}$ & $(1.31 \pm 0.49) \times 10^{-9}$ \\
H10 & X-shooter & 379.79 & 379.74 & $(1.89 \pm 0.72) \times 10^{-17}$ & $(5.47 \pm 2.17) \times 10^{-3}$ & $(1.06 \pm 0.47) \times 10^{-9}$ \\
H11 & X-shooter & 377.06 & 377.04 & $(1.58 \pm 0.68) \times 10^{-17}$ & $(5.60 \pm 2.53) \times 10^{-3}$ & $(1.09 \pm 0.54) \times 10^{-9}$ \\
Pa$\beta$ & X-shooter & 1281.81 & 1281.66 & $(2.93 \pm 0.35) \times 10^{-17}$ & $(5.91 \pm 0.74) \times 10^{-3}$ & $(1.15 \pm 0.27) \times 10^{-9}$ \\
Pa$\gamma$ & X-shooter & 1093.81 & 1093.68 & $(2.61 \pm 0.41) \times 10^{-17}$ & $(4.81 \pm 0.95) \times 10^{-3}$ & $(9.34 \pm 2.58) \times 10^{-10}$ \\
Pa$\delta$ & X-shooter & 1004.94 & 1004.82 & $(1.99 \pm 0.49) \times 10^{-17}$ & $(6.21 \pm 1.87) \times 10^{-3}$ & $(1.21 \pm 0.43) \times 10^{-9}$ \\
Pa8 & X-shooter & 954.60 & 954.42 & $(9.60 \pm 0.35) \times 10^{-18}$ & $(3.40 \pm 0.14) \times 10^{-3}$ & $(6.61 \pm 1.31) \times 10^{-10}$ \\
Pa9 & X-shooter & 922.90 & 922.78 & $(1.05 \pm 0.23) \times 10^{-17}$ & $(4.28 \pm 1.09) \times 10^{-3}$ & $(8.31 \pm 2.65) \times 10^{-10}$ \\
Pa10 & X-shooter & 901.49 & 901.32 & $(6.49 \pm 1.17) \times 10^{-18}$ & $(2.72 \pm 0.56) \times 10^{-3}$ & $(5.28 \pm 1.50) \times 10^{-10}$ \\
Br$\gamma$ & X-shooter & 2166.12 & 2165.11 & $(5.61 \pm 1.53) \times 10^{-18}$ & $(3.68 \pm 1.19) \times 10^{-3}$ & $(7.13 \pm 2.70) \times 10^{-10}$ \\
He I & X-shooter & 447.15 & 447.10 & $(5.60 \pm 1.81) \times 10^{-18}$ & $(5.89 \pm 2.02) \times 10^{-3}$ & $(1.14 \pm 0.45) \times 10^{-9}$ \\
He I $+$ Fe I & X-shooter & 492.19 & 492.34 & $(8.91 \pm 1.99) \times 10^{-18}$ & $(1.03 \pm 0.22) \times 10^{-2}$ & $(2.00 \pm 0.58) \times 10^{-9}$ \\
He I & X-shooter & 587.56 & 587.50 & $(1.21 \pm 0.18) \times 10^{-17}$ & $(6.53 \pm 1.14) \times 10^{-3}$ & $(1.27 \pm 0.33) \times 10^{-9}$ \\
He I & X-shooter & 667.82 & 667.74 & $(6.60 \pm 1.36) \times 10^{-18}$ & $(1.02 \pm 0.26) \times 10^{-2}$ & $(1.98 \pm 0.64) \times 10^{-9}$ \\
He I & X-shooter & 706.52 & 706.44 & $(4.20 \pm 0.79) \times 10^{-18}$ & $(8.34 \pm 1.85) \times 10^{-3}$ & $(1.62 \pm 0.48) \times 10^{-9}$ \\
Ca II (K) & X-shooter & 393.37 & 393.34 & $(3.97 \pm 0.46) \times 10^{-17}$ & $(6.14 \pm 0.74) \times 10^{-3}$ & $(1.19 \pm 0.27) \times 10^{-9}$ \\
Ca II & X-shooter & 849.80 & 849.70 & $(3.82 \pm 0.17) \times 10^{-17}$ & $(1.13 \pm 0.05) \times 10^{-2}$ & $(2.20 \pm 0.44) \times 10^{-9}$ \\
Ca II & X-shooter & 854.21 & 854.12 & $(4.93 \pm 0.19) \times 10^{-17}$ & $(1.21 \pm 0.05) \times 10^{-2}$ & $(2.35 \pm 0.47) \times 10^{-9}$ \\
Ca II & X-shooter & 866.21 & 866.12 & $(4.25 \pm 0.19) \times 10^{-17}$ & $(1.19 \pm 0.05) \times 10^{-2}$ & $(2.30 \pm 0.46) \times 10^{-9}$ \\
O I & X-shooter & 777.31 & 777.10 & $(5.89 \pm 1.22) \times 10^{-18}$ & $(6.29 \pm 1.66) \times 10^{-3}$ & $(1.22 \pm 0.40) \times 10^{-9}$ \\
O I & X-shooter & 844.64 & 844.54 & $(8.71 \pm 1.27) \times 10^{-18}$ & $(6.43 \pm 1.01) \times 10^{-3}$ & $(1.25 \pm 0.31) \times 10^{-9}$ \\
Pa$\beta$ & IRTF & 1281.81 & 1282.11 & $(4.87 \pm 0.03) \times 10^{-17}$ & $(1.01 \pm 0.01) \times 10^{-2}$ & $(1.96 \pm 0.38) \times 10^{-9}$ \\
Br$\gamma$ & IRTF & 2166.12 & 2166.51 & $(8.75 \pm 0.20) \times 10^{-18}$ & $(6.24 \pm 0.18) \times 10^{-3}$ & $(1.21 \pm 0.24) \times 10^{-9}$ \\
Ca II & IRTF & 849.80 & 850.06 & $(1.48 \pm 0.03) \times 10^{-17}$ & $(4.45 \pm 0.11) \times 10^{-3}$ & $(8.63 \pm 1.69) \times 10^{-10}$ \\
Ca II & IRTF & 854.21 & 854.54 & $(1.91 \pm 0.03) \times 10^{-17}$ & $(4.85 \pm 0.08) \times 10^{-3}$ & $(9.41 \pm 1.84) \times 10^{-10}$ \\
Ca II & IRTF & 866.21 & 866.59 & $(1.85 \pm 0.03) \times 10^{-17}$ & $(5.46 \pm 0.10) \times 10^{-3}$ & $(1.06 \pm 0.21) \times 10^{-9}$ \\
Pa$\gamma$ & IRTF & 1093.81 & 1094.14 & $(3.78 \pm 0.03) \times 10^{-17}$ & $(7.62 \pm 0.12) \times 10^{-3}$ & $(1.48 \pm 0.29) \times 10^{-9}$ \\
Pa$\delta$ & IRTF & 1004.94 & 1005.13 & $(2.91 \pm 0.03) \times 10^{-17}$ & $(9.90 \pm 0.15) \times 10^{-3}$ & $(1.92 \pm 0.37) \times 10^{-9}$ \\
Pa8 & IRTF & 954.60 & 954.79 & $(2.32 \pm 0.04) \times 10^{-17}$ & $(8.93 \pm 0.17) \times 10^{-3}$ & $(1.73 \pm 0.34) \times 10^{-9}$ \\
Pa9 & IRTF & 922.90 & 923.09 & $(1.36 \pm 0.03) \times 10^{-17}$ & $(5.79 \pm 0.18) \times 10^{-3}$ & $(1.12 \pm 0.22) \times 10^{-9}$ \\
Pa10 & IRTF & 901.49 & 901.72 & $(9.94 \pm 0.29) \times 10^{-18}$ & $(4.45 \pm 0.16) \times 10^{-3}$ & $(8.63 \pm 1.71) \times 10^{-10}$ \\
\hline
\end{tabular}
\end{table*}

\begin{table*}[h!]
\centering
\caption{Summary of the line fluxes, accretion luminosities, and mass accretion rates during quiescence.}
\label{tab:fluxes_accrates_q}
\begin{tabular}{llrrlll}
\toprule
Line & Telescope & $\lambda_{\rm{lab}}$ (nm) & $\lambda$ (nm) & $F$ (W/m$^2$) & $L_{\rm{acc}}$ ($L_\odot$) & $\dot{M}$ ($M_\odot$/yr) \\
\midrule
Br$\gamma$ & IRTF (2025.03.) & 2166.12 & 2165.11 & $(2.97 \pm 0.19) \times 10^{-18}$ & $(1.73 \pm 0.13) \times 10^{-3}$ & $(3.35 \pm 0.70) \times 10^{-10}$ \\
Pa$\beta$ & IRTF (2024.08.) & 1281.81 & 1281.69 & $(3.45 \pm 0.28) \times 10^{-18}$ & $(6.11 \pm 0.52) \times 10^{-4}$ & $(1.19 \pm 0.25) \times 10^{-10}$ \\
Br$\gamma$ & IRTF (2024.08.) & 2166.12 & 2167.97 & $(2.32 \pm 0.13) \times 10^{-18}$ & $(1.29 \pm 0.08) \times 10^{-3}$ & $(2.50 \pm 0.51) \times 10^{-10}$ \\
Pa$\gamma$ & IRTF (2024.08.) & 1093.81 & 1094.33 & $(1.96 \pm 0.23) \times 10^{-18}$ & $(1.94 \pm 0.28) \times 10^{-4}$ & $(3.77 \pm 0.92) \times 10^{-11}$ \\
\bottomrule
\end{tabular}
\end{table*}

%

%
%

\end{document}